\documentstyle[12pt]{article}
\newcommand{\be}{\begin{equation}}
\newcommand{\bea}{\begin{eqnarray}}
 
\newcommand{\eea}{\end{eqnarray}}
\newcommand{\ba}{\begin{array}}
\newcommand{\ea}{\end{array}}
\newcommand{\ee}{\end{equation}}

\expandafter\ifx\csname mathbbm\endcsname\relax

\else

\fi
\begin{document}
\begin{titlepage}
\hfill
\vbox{
    \halign{#\hfil         \cr
           hep-th/0208187 \cr
           IPM/P-2002/034 \cr
           } 
      }  
\vspace*{20mm}
\begin{center}
{\Large {\bf Branes in Time-Dependent Backgrounds and AdS/CFT
Correspondence}\\ }

\vspace*{15mm} \vspace*{1mm} {Mohsen Alishahiha\footnote{
e-mail: alishah@ipm.ir} and Shahrokh Parvizi\footnote{e-mail:
parvizi@theory.ipm.ac.ir} }

\vspace*{1cm}

{\it Institute for Studies in Theoretical Physics and Mathematics (IPM)\\
P.O. Box 19395-5531, Tehran, Iran\\}

\vspace*{1cm}

\end{center}

\begin{abstract}
We study supergravity solutions of Dp-branes in the time-dependent orbifold
background. We show that worldvolume theories decouple from the bulk
gravity for $p$ less than six. Along AdS/CFT correspondence, these solutions 
could provide the gravity 
description of noncommutative field theory with time-dependent
noncommutative parameter. Type II NS5-brane (M5-brane) in the presence of 
RR $n$-form for $n=0,\cdots, 4$ (C field) in this time-dependent background 
have also been studied.
\end{abstract}

\end{titlepage}

\section{Introduction}
Recently string theory in  time-dependent backgrounds has attracted much 
interest (recent papers on this subject include \cite{HS1}-\cite{Sim}). 
This is because in order to apply the string theory in cosmology we would need
to know the string theory in time-dependent backgrounds. As an example one
can consider orbifold quotients of Minkowski space-time which non-trivially
act on both space and time \cite{{HS1},{FS},{KOSST},{BHKN},{CC},{Nek},
{LMS1},{CCK},{LMS2},{FMc},{HS},{Sim}}. We note, however, that in general
orbifolds of space-time leads to singularities corresponding to the closed
light-like loops. Nevertheless one can resolve these singularities by making 
use of the null-brane backgrounds \cite{{FS},{LMS2},{FMc},{Sim}} which is
obtained form the space-time orbifold together with an extra shift.
In the next section we review this geometry.

This null-brane background preserves one-half of the original flat space
supersymmetry. It has also a null Killing vector which makes it possible
to use light-cone quantization. The stability of this geometry has also
been studied in \cite{{Law},{FMc},{HPo}}.

Open string sector in this background has  recently been considered
\cite{{HS},{Sim}}. 
This, for example, can be done by using D-branes 
for probing the geometry. D3-brane
probing this background has been studied in \cite{HS} where the
authors argued that the D3-brane worldvolume decouples from the
bulk gravity leading to a noncommutative gauge theory with 
time-dependent noncommutativity parameter. 
This is the aim of this article to generalize this consideration for
other D-branes as well as NS5-brane. In the NS5-brane case the situation 
reminisces the ODp-theory, namely we will find NS5-brane solution in
the presence of RR $n$-form for $n=0,\cdots,4$ in the time-dependent
background. We shall also consider M5-brane in an 11-dimensional 
time-dependent background which can be obtained in the same way as 
the one in string theory. Namely we could start from an 11-dimensional
null-brane geometry of M-theory. The theory on the corresponding
M5-brane  can be thought as a new deformation of (0,2) theory. This
system is very similar to OM-theory \cite{GMSS}.

The organization of the paper is as follows. In section 2 we will
review the null-brane background and then we shall probe this background
by Dp-brane for $2<p\leq 6$. We will argue that the worldvolume theory 
decouples from gravity for $p\leq 5$ and therefore this background would
provide a gravity description for noncommutative field theory with 
time-dependent noncommutativity parameter. In section 3 we will consider
the near horizon limit of the background obtained in section 2. We shall
briefly study the phase structure of the theory as well. In section
4 type II NS5-branes in the null-brane geometry are considered which lead
to new deformation of the little string theory. In section 5 M5-brane 
probing 11-dimensional null-brane geometry is studied. Section 6 is
devoted to the comments and conclusions.    

\section{Supergravity solution}
In this section we shall briefly review the structure of null-brane
geometry. Then in order to study the properties of this background in 
string theory we will probe the geometry by D-branes. This background 
could change the worldvolume theory of the brane leading, probably, 
to a new field theory.

We start with a 10-dimensional flat spacetime $R^{1,9}$ with metric 
\be
ds^2=-2dx^+dx^-+dx^2+dz^2+d{\vec Y}_6^2\;,
\ee
where $d{\vec Y}_6^2$ is the flat 6-dimensional space. Now consider
the orbifold obtained by identifying \cite{{LMS1},{LMS2}}
\be
\pmatrix{x^+\cr x\cr x^-\cr z}\sim\pmatrix{x^+\cr x+2\pi n x^+\cr
x^-+2\pi n x+2(\pi n)^2x^+\cr z+2\pi n \beta},\;\;\;\;\;\;\;\;n\in Z\;,
\ee
while leaves  $d{\vec Y}_6^2$ unchanged and $\beta$ is a constant.
 The above orbifold background is 
called the null-brane \cite{FS}.  

It is useful to introduce a new set of coordinates as following
\be
x^+=y^+,\;\;\;\;\;x=y^+y,\;\;\;\;\;x^-=y^-+{1\over 2}y^+y^2\;,
\ee 
and other coordinates remain the same as before. In this coordinates the
orbifold metric reads
\be
ds^2=-2dy^+dy^-+(y^+)^2dy^2+dz^2+d{\vec Y}_6^2\;,
\label{BAC}
\ee
and the orbifold identification becomes
\be
\pmatrix{y^+\cr y\cr y^-\cr z}\sim\pmatrix{y^+\cr y+2\pi n \cr
y^- \cr z+2\pi n \beta}\;.
\ee

Now we want to probe this background with a system of $N$ 
Dp-branes for $p>2$ with worldvolume directions $(y^+,y^-,y,z,
x_1\cdots x_{p-3})$ where $x_i$'s are $(p-3)$-dimensional 
subspace of ${\vec Y}_6$.
The procedure is very similar to that in \cite{BDGKR} (see also
\cite{AY}) to construct the supergravity dual of the noncommutative
dipole theory. To do this we start with the flat Dp-brane in the
$y$ coordinates
\bea
ds^2&=&f^{-1/2}\bigg{(}-2dy^+dy^-+(y^+)^2dy^2+dz^2+\sum_{i=1}^{p-3}dx_i^2
\bigg{)}+
f^{1/2}(dr^2+r^2d\Omega^2_{8-p})\;,\cr&&\cr
e^{2\phi}&=&f^{(3-p)/2},\;\;\;\;\;dC_{+-yz1\cdots(p-3)}\sim\partial_rf^{-1},
\;\;\;\;\;f=1+{R^{7-p}\over r^{7-p}}\;, 
\eea
where $R^{7-p}=c_p g_sNl_s^{7-p}$ with $c_p= 2^{5-p} 
\pi^{\frac{5-p}{2}} \Gamma (\frac{7-p}{2})$. Here
$z$ is compactified on a circle of radius $\beta$. T-dualizing along $z$ 
coordinate leads to the D(p-1)-brane smeared over one direction
\bea
ds^2&=&f^{-1/2}\bigg{(}-2dy^+dy^-+(y^+)^2dy^2+\sum_{i=1}^{p-3}dx_i^2
\bigg{)}+
f^{1/2}(dz^2+dr^2+r^2d\Omega^2_{8-p})\;,\cr&&\cr
e^{2\phi}&=&f^{(4-p)/2},\;\;\;\;\;dC_{+-y1\cdots(p-3)}\sim\partial_rf^{-1},
\;\;\;\;\;f=1+{R^{7-p}\over r^{7-p}}\;. 
\label{TDU}
\eea
We now twist $y$ coordinate so that under identification of $z$ one gets
\be
z\sim z+2\pi \beta,\;\;\;\;\;y\sim y+2\pi\;.
\ee
One can define a new coordinate in which $y\rightarrow y-{z\over \beta}$ 
such that the identification becomes trivial
\be
z\sim z+2\pi \beta,\;\;\;\;\;y\sim y\;.
\ee
In this new coordinates the T-dualized metric (\ref{TDU}) reads
\bea
ds^2&=&f^{-1/2}\bigg{(}-2dy^+dy^-+\sum_{i=1}^{p-3}dx_i^2\bigg{)}
+f^{1/2}(dr^2+r^2d\Omega^2_{8-p})\cr &&\cr
&+& f^{-1/2}(y^+)^2dy^2+f^{1/2}(1+{(y^+)^2\over \beta^2 f})dz^2
+2f^{-1/2}{(y^+)^2\over \beta}dydz\;,
\eea
T-dualizing back along $z$ direction and using the T-duality rules
\cite{T-D} we will find a supergravity solution of Dp-brane probing 
the orbifold background (\ref{BAC}) as following\footnote{The 
supersymmetry of branes in the null-brane
geometry in string theory and M-theory was also studied in 
\cite{{Ofa1},{Ofa2}}. In fact the solution we have found here is
dual to one of the possible reductions of the delocalised 
M2-brane discussed in section 3.2 of paper \cite{Ofa1}. 
This solution is the special case of the reduction 
labelled (A) in that section where the $\theta_i = 0$.  
We would like to thank J. Figueroa-O'Farrill for bringing
our attention to this point.}
\bea
ds^2&=&f^{-1/2}\bigg{(}-2dy^+dy^-+h((y^+)^2dy^2+dz^2)+
\sum_{i=1}^{p-3}dx_i^2\bigg{)}+
f^{1/2}(dr^2+r^2d\Omega^2_{8-p})\;,\cr&&\cr
e^{2\phi}&=&h f^{(3-p)/2},\;\;\;\;\;\;\;\;\;h^{-1}=1+{(y^+)^2\over \beta^2 f},
\;\;\;\;\;f=1+{R^{7-p}\over r^{7-p}}\;,\cr &&\cr 
B_{yz}&=&{(y^+)^2\over \beta}hf^{-1}\;,\;\;\;\;\;dC_{+-yz1\cdots(p-3)}
\sim\partial_rf^{-1},
\label{BACKGROUND}
\eea

In general given a supergravity solution of a system 
of branes, it is not clear whether the solution would give a 
well-defined description of some field theory. In fact, we must 
check and see whether there is a well-defined field theory on the 
brane worldvolume  which decouples from the bulk gravity. To see 
this, one might calculate scattering amplitude for gravitons 
\cite{{GKT},{GK}} which can be done by computing the gravitons 
absorption cross section. If there is a limit (decoupling limit) 
where the gravitons absorption cross section vanishes, we have a 
field theory which decouples from the gravity.
Using the same method as in \cite{AIO} one can see that the background
(\ref{BACKGROUND}) give a well-defined theory in a decoupling limit for
$p<6$. 

More precisely, let us perturb the metric of the background 
(\ref{BACKGROUND}) by
\be 
g_{\mu\nu} = \bar{g}_{\mu\nu} + h_{\mu\nu} \hspace{1cm} \mu,\nu=0,\cdots,9, 
\ee
where by $\bar{g}_{\mu\nu}$ we denote the background metric 
(\ref{BACKGROUND})
and $h_{\mu\nu}$ is the perturbation. We consider s-wave gravitons as following
\be
h_{\mu\nu} = \epsilon_{\mu\nu} \Psi(r,y^+,y^-) , \hspace{1cm} 
\mu = 0,\cdots, p \; .
\ee
Using the linearized equations of motion 
of type II supergravity we find following equation for transverse 
gravitons
\be
\partial_{\mu} \left(\sqrt{-g} e^{- 2 \phi} g^{\mu\nu} 
\partial_{\nu}\Psi\right)=0,
\ee
which leads to
\be
\partial_r(r^{(8-p)}\partial_r\Psi)-2fr^{8-p}\partial_+\partial_-
\Psi-{r^{8-p}\over y^+}f\partial_-\Psi=0\;.
\ee
As an ansatz we take $\Psi(r,y^+,y^-)=h(r)Y(y^+,y^-)$. 
Thus we find two differential equations 
\be
\partial_r^2h+{8-p\over r}\partial_rh+\omega^2f
h=0,\;\;\;\;\;
2\partial_+\partial_-Y+{1\over y^+}\partial_-Y-\omega^2Y=0,
\ee
The differential equation for $Y$ can be easily solved 
\be
Y={1\over \sqrt{ y^+}}\;e^{{i\omega\over \sqrt 2}(y^++y^-)}\;.
\ee
This time-dependence form of the solution reflects the fact that 
at large $r$ the space is not flat. From the radial equation one can 
read the potential by writing it in the form of a Schr\"odinger-like 
equation. Doing so we get the  following Schr\"odinger-like equation
\be
\partial_{\rho}^2 \varphi(\rho) +V_p(\rho) \varphi(\rho)=0\;,
\ee 
where 
\be V_p(\rho) = -\left(1+\frac{c_pNg_s(\omega l_s)^{7-p}}
{\rho^{7-p}}\right)+ \frac{(8-p)(6-p)}{4 \rho^2}
\ee 
with $\rho=\omega r$.

This potential is the same as the one  found in \cite{AIO} for the 
ordinary D-branes as well as branes in the presence of B field. 
Therefore we conclude that we have a decoupled theory 
living on the worldvolume of Dp-brane for $p\leq 5$.

\section{Decoupling limit}

In this section we study the near horizon limit of the supergravity solution 
(\ref{BACKGROUND}). Having a decoupling limit, this could provide a gravity
description of the theory along the AdS/CFT correspondence \cite{MAL}.

The decoupling limit of the background (\ref{BACKGROUND}) is defined as
a limit in which $l_s\rightarrow 0$ while keeping the following
quantities fixed \cite{HS}
\be
u={r\over l_s^2},\;\;\;\;\;b={\beta\over l_s^2},\;\;\;\;\;
{\tilde g}=g_sl_s^{p-3}\;.
\ee
In this limit the supergravity solution (\ref{BACKGROUND}) reads
\bea
l_s^{-2}ds^2&=&{u^{(7-p)/2}\over \sqrt{g_{YM}^2N}}
\bigg{(}-2dy^+dy^-+h((y^+)^2dy^2+dz^2)+
\sum_{i=1}^{p-3}dx_i^2\bigg{)}\cr &&\cr
&+&{\sqrt{g_{YM}^2N}\over u^{(7-p)/2}}(du^2+u^2d\Omega^2_{8-p})\;,
\cr&&\cr e^{2\phi}&=&{\tilde g}^2 
\left({g_{YM}^2N\over u^{7-p}}\right)^{{3-p\over 2}}h,
\;\;\;\;\; B_{yz}={l_s^2\over b}\;
{(y^+)^2u^{7-p}\over g_{YM}^2N}h\;,
\label{NEAR}
\eea
where $h^{-1}=1+{(y^+)^2u^{7-p}\over g_{YM}^2b^2N}$ 
and $g_{YM}^2=c_p{\tilde g}$.

Following \cite{HS} string theory on these backgrounds provides the
gravity description of noncommutative gauge theories with nonconstant
noncommutativity parameter in various dimensions. 

The effective dimensionless coupling constant in the corresponding 
noncommutative field theory can be defined as 
following \cite{IMSY}
\be
g_{\rm eff}^2\sim g_{\rm YM}^2 N u^{p-3}\; .
\ee
The scalar curvature of the metric in eq. (\ref{NEAR}) has the 
behavior
\be
l_s^2{\cal R}\sim \frac{1}{g_{\rm eff}}\; .
\ee
Thus the perturbative calculation in 
noncommutative field theory can be trusted when $g_{\rm eff}\ll 1$, 
while when  $g_{\rm eff}\gg 1$ the supergravity description is valid. 
We note also that the expression for dilaton in (\ref{NEAR}) 
can be recast to 
\be
e^{\phi}={1\over N}\;g_{\rm eff}^{(7-p)/4}h^{1/2}\;.
\label{DIL}
\ee
Keeping $g_{\rm eff}$ and $h$ fixed we see from 
(\ref{DIL}) that $e^{\phi}\sim 1/N$. Therefore the string
loop expansion corresponds to $1/N$ expansion of noncommutative gauge 
theory.

Since the scalar curvature is time-independent,
as far as the effective gauge coupling is concerned,
the situation is the same as ordinary brane solution. 
But since the dilaton is time-dependent this will change the phase structure
of the theory. In particular at given fixed energy the effective string 
coupling will change with time. There is a critical time 
$y^+_c=g_{YM}b\sqrt{N}/u^{(7-p)/2}$ where for $y^+>y_c^+$ the 
noncommutative effects become important and the effective string
coupling becomes
\be
e^{2\phi}\sim {b (g_{YM}^2N)^{(9-p)/2}\over N}\;{u^{(p-5)(7-p)/2}\over 
(y^{+})^2}\;.
\ee
Therefore the effective string coupling decreases in time, and 
thus under time evolution the gravity description becomes 
more applicable.

On the other hand at any given time the phase structure of the 
theory is the same as that in noncommutative field theory with 
constant noncommutativity parameter, namely those studied in 
\cite{MOS} (see also \cite{allpapers}). The only difference is
that the distinguished points where the description of the theory
has to be changed is now time-dependent.

In the notation of \cite{MOS} the dimensionless effective 
noncommutative parameter is given by
\be
a^{\rm eff}= \left({y^+ u^2\over b g_{\rm eff}}\right)^{2\over 7-p}
\;.
\label{effnon}
\ee
At large distances
$L\gg \sqrt{y^+/bg_{\rm eff}}$ the noncommutative effects are small
and the effective description of the worldvolume theory is in terms
of a commutative field theory. Note that this distance is time-dependent
which means that under time evolution of the theory the distance where the
noncommutative effects are negligible becomes very large. 

Form the expression of the  dimensionless effective noncommutative 
parameter (\ref{effnon}) one can read the noncommutative parameter  
seen by the gauge theory. In fact we get 
\be
[y,z]\sim {y^+\over b},
\ee
which is in agreement with the field theory consideration \cite{HS}.

\section{NS5-brane in the presence of RR field}

The worldvolume theory of NS5-branes in the presence of electric
RR field decouples from bulk gravity leading to an interacting
theory with light open Dp-brane. These theories are known as
open Dp-brane theory or in short ODp-theory 
\cite{{GMMS},{HARM1}}. The supergravity description of these theories
have been studied in \cite{{all},{ALI1},{AOR},{Mitra}}.

In this section we shall study time-dependent background of 
type II NS5-brane in the presence of RR field. To do this, we 
start from D5-brane solution presented in the previous section. 
By making use of S-duality one can find the type IIB NS 5-brane
in the presence of RR 2-form \footnote{Under S-duality we have
$\phi\rightarrow -\phi$ and $ds^2\rightarrow e^{-\phi}ds^2$ where
$ds^2$ is the metric in string frame. Moreover NS B field gets 
change to the RR 2-form.}. Then a series of T-dualities will
generate other possible RR fields. In fact by this procedure one
gets 
\bea
ds^2&=&h^{-1/2}\bigg{[}-2dy^+dy^-+h((y^+)^2dy^2+\sum_{i=1}^{n}dx_i^2
)+\sum_{j=n+1}^3dx_j^2+f(dr^2+r^2d\Omega_3^2)\bigg{]},\cr &&\cr
e^{2\phi}&=&fh^{(n-3)/2},\;\;\;\;\;\;\;C^{(n+1)}_{y1\cdots n}=
{1\over g_s}{(y^+)^2\over \beta}hf^{-1},\;\;\;\;\;
f=1+{ N l_s^2\over r^2}\;,
\label{NS1}
\eea
which is type II NS5-brane in the presence of RR $(n+1)$-form 
for $n=0,1,2,3$.

We note, however, that in the D5-brane solution (\ref{BACKGROUND}) the
metric is not symmetric under exchanging of 
$y\leftrightarrow z$.\footnote{Note that this is not the case for the 
noncummutative field theory with constant noncommutativity parameter.}
Therefore we have another choice of doing the series of T-dualities in which
we have also make a T-duality along $y$ direction. Doing so we get
another solution  of type II NS5-brane in the presence of different possible
RR fields as following
\bea
ds^2&=&h^{-1/2}\bigg{[}-2dy^+dy^-+{dy^2\over (y^+)^2}+h
\sum_{i=1}^{n}dx_i^2+\sum_{j=n+1}^3dx_j^2+f(dr^2+r^2d\Omega_3^2)
\bigg{]},\cr &&\cr
e^{2\phi}&=&fh^{(n-4)/2},\;\;\;\;\;\;\;C^{(n)}_{1\cdots n}={1\over g_s}
{(y^+)^2\over \beta}hf^{-1},
\;\;\;\;\;f=1+{ N l_s^2\over r^2}\;,
\label{NS2}
\eea
which is the type II NS5-brane in the presence of RR $n$-form for $n=0,1,2,3$.

Having a supergravity solution one could proceed to consider the decoupling
limit of the theory. The decoupling limit of the NS5-brane can be 
obtained from D5-brane using S-duality in which 
$l_s^2\rightarrow g_sl_s^2$ and $g_s\rightarrow g_{s}^{-1}$. Thus the
decoupling limit is given by a limit in which $g_s\rightarrow 0$ while 
keeping the following quantities fixed\footnote{To make $u$ of dimension of
energy, we have also added $l_s$ in the definition of $u$, 
similarly for $b$.} 
\be
u={r\over g_sl_s^2},\;\;\;\;\;b={\beta\over g_sl_s^2},\;\;\;\;\;
l_s={\rm fixed}\;,
\ee
which is very similar to the decoupling limit of the little
string theory \cite{Seiberg} (see also \cite{BER}). In this limit the 
supergravity solutions 
(\ref{NS1}) and (\ref{NS2}) read
\bea
ds^2&=&h^{-1/2}\bigg{[}-2dy^+dy^-+h((y^+)^2dy^2+\sum_{i=1}^{n}dx_i^2
)+\sum_{j=n+1}^3dx_j^2\cr &&\cr
&&\;\;\;\;\;\;\;\;\;\;\;+{ Nl_s^2\over u^2}
(du^2+u^2d\Omega_3^2)\bigg{]},\cr &&\cr
e^{2\phi}&=&{ Nl_s^2\over u^2}h^{(n-3)/2},\;\;\;\;\;\;\;
C^{(n+1)}_{y1\cdots n}=
{l_s^2\over b}{(y^+)^2u^2\over  N }h\;,
\eea
for (\ref{NS1}), and 
\bea
ds^2&=&h^{-1/2}\bigg{[}-2dy^+dy^-+{dy^2\over (y^+)^2}+h
\sum_{i=1}^{n}dx_i^2+\sum_{j=n+1}^3dx_j^2+{ Nl_s^2\over u^2}
(du^2+u^2d\Omega_3^2)
\bigg{]},\cr &&\cr
e^{2\phi}&=&{ Nl_s^2\over u^2}h^{(n-4)/2},\;\;\;\;\;\;\;
C^{(n)}_{1\cdots n}={l_s^2\over b}{(y^+)^2u^2\over  N }h \;,
\eea
for (\ref{NS2}). Here $h^{-1}=1+(y^+)^2u^2/ N b^2$.

The curvature of the above supergravity solutions reads
\be
l_s^2{\cal R}\sim {1\over (1+{(y^+)^2u^2\over b^2N})^{1/2}}
\ee
therefore for large $u$ the curvature is small and the supergravity
solution provides a good description of the theory.

The same as previous section one can consider scalar field
scattering of the NS5-brane. The corresponding wave equation for an 
ansatz of the form $\Psi(r,y^+,y^-)={\varphi(r)\over \sqrt{y^+}}\;
e^{{i\omega\over \sqrt{2}}(y^++Y^-)}$ leads to the following  
Schr\"odinger-like equation
\be
\partial_{\rho}^2 \varphi(\rho) +V_p(\rho) \varphi(\rho)=0\;,
\ee 
where 
\be V_p(\rho) = -1+({3\over 4}-1)\frac{N\omega^2 l^2_s}
{\rho^2}\;,
\ee 
with $\rho=\omega r$. Following \cite{MS} we conclude that the theory
has a mass gap of order $m_{\rm gap}\sim 1/\sqrt{Nl_s^2}$. 
This is exactly the same mass gap as the one in the little string theory,
namely the presence of RR field has not changed the mass gap. Note  
that in the ODp-theory the mass gap will change because of the
presence of RR field. In fact in this case the mass gap is given by
noncommutative effective parameter which is given  by the value of
electric RR field \cite{ALI1}. In this sense these theories are closer
to the little string theory than ODp-theory.

\section{M-theory 5-brane}

The supergravity solution of M5-brane in the presence of nonzero 
$C$ 3-form along the worldvolume of the brane has been considered in 
\cite{MOS}. This worldvolume theory in the decoupling limit is a
well-defined quantum theory with light open membrane, known as
OM-theory \cite{GMSS}. M5-brane solution in the presence of C 
field with
two legs along the worldvolume directions has also been studied in 
\cite{AY} in the context of noncommutative ``{\it discpole}'' 
theory \cite{DGR}.

In this section we would like to study the M5-brane probing an
11-dimensional time-dependent orbifold background similar to
(\ref{BAC}). To find the supergravity solution, one can start from
D4-brane solution and then lift it up to the 11-dimensional 
supergravity. By making use of the relation between 11-dimensional 
supergravity solution and its corresponding 10-dimensional 
dimensional solution 
\be
ds_{11}^2=e^{{4\phi\over 3}}(dx_{11}+C_{\mu}dx^{\mu})^2+
e^{-{2\phi \over 3}}ds_{10}^2
\ee 
one can lift up the supergravity solution of D4-brane in (\ref{BACKGROUND}),
to find
\bea
ds^2&=&(hf)^{-1/3}\bigg{[}-2dy^+dy^-+h((y^+)^2dy^2+dx_1^2+dx_2^2)
+dx_3^2+f(dr^2+r^2d\Omega_4^2)\bigg{]},\cr
C_{y12}&=&{(y^+)^2\over \beta}hf^{-1},\;\;\;\;\;f=1+{\pi N l_p^3\over 
r^3}\;,\;\;\;\;\;h^{-1}=1+{(y^+)^2\over \beta^2}\;f^{-1}\;.
\label{M5}
\eea
where $l_p$ is 11-dimensional Plank length.
Of course we have another $C$ form representing the charge of $N$ 
M5-branes. We note also that under reduction to 10-dimensional solution
along a transverse direction we will get the type IIA NS5-brane
solution in the presence of RR 3-form as in (\ref{NS1}).

The decoupling limit of the theory is defined by a limit in which
$l_p\rightarrow 0$ while keeping the following quantities fixed
\be
u^2={r\over l_p^3},\;\;\;\;\;\;\;b^2={\beta\over l_p^3}\;.
\ee
In this limit the supergravity solution (\ref{M5}) reads
\bea
l_p^{-2}ds^2&=&{u^2\over (\pi N)^{1/3}}h^{-1/3}
\bigg{[}-2dy^+dy^-+h((y^+)^2dy^2+dx_1^2+dx_2^2)
+dx_3^2\cr &&\cr
&&\;\;\;\;\;\;\;\;\;\;\;\;\;\;\;\;\;\;\;\;\;\;\;+
{\pi N\over u^4}(4du^2+u^2d\Omega_4^2)\bigg{]},\cr
C_{y12}&=&{l_p^3\over b}{(y^+)^2u^6\over \pi N }h\;,
\;\;\;\;\;h^{-1}=1+{(y^+)^2u^6\over \pi N b^4}\;.
\eea
The curvature of the above metric reads
\be
l_p^{2}{\cal R}\sim {1\over  N^{2/3}}\;{1\over 
(1+{(y^+)^2u^6\over \pi N b^4})^{1/3}}\;.
\ee
Therefore we can trust the supergravity solution in the UV for
a given time. On the other hand under time evolution
the curvature decreases and thus the good description is given
in terms of supergravity.

\section{Conclusions} 

In this paper we have studied supergravity solution of 
various type II superstring
branes in a time-dependent background. The background we have probed by
branes is null-brane geometry which is obtained by space-time orbifold
accompanied by a shift. The shift will resolve the singularities which 
could appear because of closed time-like loops.

We have seen that the worldvolume theory decouple from gravity for Dp-brane
with $p\leq 5$. Following \cite{HS} these backgrounds could provide gravity
description of noncommutative field theories with time-dependent 
noncommutativity parameter. These backgrounds preserve 8 supersymmetries 
out of 32 supersymmetries presented in the flat space. Indeed the 
time-dependent orbifold we have considered in this paper breaks one-half
of the supersymmetry. Adding brane in the theory will break one-half more. 
Therefore the noncommutative field theories we have studied in this paper
preserve only one half of the supersymmetry of the corresponding 
noncommutative field theories with constant noncommutativity parameter.
In this sense the situation is very similar to the noncommutative
dipole field theory where the  maximal possible SUSY is 8 
supercharges \cite{{BDGKR},{Jabbari}, {AY}}.

We have also considered type II NS5-branes in the null-brane background.
In the decoupling limit we have found a series of six dimensional theories
which are identified by an RR form. These theories can be thought as a new 
deformation of the little string theory. The structure of these theories  
reminisces the structure of ODp-theories \cite{GMSS}, 
though, as we have seen, the mass 
gap of theory is given by string length while in ODp theory it is fixed
by deformation parameter or the value of electric RR field at infinity.

We note also that using T-duality we have been able to find a new set of 
NS5-brane solutions in the presence of RR field (\ref{NS2}). This could 
also give another deformation of the little string theory. Now starting 
from 
gravity solution (\ref{NS2}) with $n=2$ one can use S-duality to find a new
gravity solution of D5-brane in a time-dependent background which is
different from one in (\ref{BACKGROUND}). Now using T-duality we will get
new Dp-brane solution in the time dependent background. It would be 
interesting to study this background as well.

Having a gravity description of a theory it is natural to compute the
Wilson loop of corresponding field theory using its gravity 
description \cite{{MAL11},{REY}}. It would therefore be interesting to 
compute the Wilson loop
in the noncommutative gauge theory studied in this paper using its gravity
dual (\ref{NEAR}).

\vspace*{.4cm}

{\bf Acknowledgments}

We would like to thank M. M. Sheikh-Jabbari for comments and discussions.


\begin{thebibliography}{99}


\bibitem{HS1}
G.T. Horowitz and A.R. Steif, ``Singular String Solutions
With Nonsingular Initial Data,'' {\em Phys.Lett.}
{\bf B258} (1991) 91. 


\bibitem{Fig}
J. Figueroa-O'Farrill, ``Breaking the M-waves,''
{\em Class.Quant.Grav.} {\bf 17} (2000) 2925,
hep-th/9904124. 

\bibitem{FS}
J. Figueroa-O'Farrill and J. Simon, ``Generalized Supersymmetric 
Fluxbranes,''{\em JHEP} {\bf 0112} (2001) 011, hep-th/0110170.

J. Simon, ``The Geometry of Null Rotation Identifications,'' 
hep-th/0203112.

\bibitem{KOSST}
J. Khoury, B. A. Ovrut, N. Seiberg, P. J. Steinhardt and N. Turok,
``From Big Crunch to Big Bang,'' {\em Phys.Rev.} {\bf D65}
(2002) 086007, hep-th/0108187.

N. Seiberg, ``From Big Crunch To Big Bang - Is It Possible?,''
hep-th/0201039.

\bibitem{BHKN}
V. Balasubramanian, S. F. Hassan, E. Keski-Vakkuri and A. Naqvi, ``
A Space-Time Orbifold: A Toy Model for a Cosmological Singularity,''
hep-th/0202187.

\bibitem{GS}
M. Gutperle and A. Strominger, ``Spacelike Branes,''
{\em JHEP} {\bf 0204} (2002) 018,
hep-th/0202210. 

\bibitem{CC}
L. Cornalba and M. S. Costa, ``A New Cosmological Scenario in 
String Theory,'' hep-th/0203031. 

\bibitem{Nek}
N. A. Nekrasov, ``Milne Universe, Tachyons, and Quantum Group,''
hep-th/0203112.

\bibitem{Sen}
A. Sen, ``Rolling Tachyon,'' {\em JHEP} {\bf 0204} (2002) 048,
hep-th/0203211. 

\bibitem{KP}
E. Kiritsis and B. Pioline, ``Strings in homogeneous gravitational 
waves and null holography,'' hep-th/0204004. 


\bibitem{TT}
A. J. Tolley and N. Turok, ``Quantum Fields in a Big Crunch/Big 
Bang Spacetime,''hep-th/0204091. 

\bibitem{Myers}
M. Kruczenski, R.C. Myers, A.W. Peet, ``Supergravity S-Branes,''
{\em JHEP} {\bf 0205} (2002) 039,
hep-th/0204144.

\bibitem{AFHS}
O. Aharony, M. Fabinger, G. Horowitz and E. Silverstein, ``
Clean Time-Dependent String Backgrounds from Bubble Baths,''
{\em JHEP} {\bf 0207} (2002) 007, hep-th/0204158. 
 
\bibitem{LMS1}
H. Liu, G. Moore and N. Seiberg, ``Strings in a Time-Dependent Orbifold,''
hep-th/0204168.

\bibitem{EGKR}
S. Elitzur, A. Giveon, D. Kutasov and E. Rabinovici, ``
From Big Bang to Big Crunch and Beyond,'' {\em JHEP}
{\bf 0206} (2002) 017,
hep-th/0204189. 

\bibitem{CCK}
L. Cornalba, M.S. Costa and C. Kounnas, ``A Resolution of the 
Cosmological Singularity with Orientifolds,''
{\em Nucl.Phys.} {\bf B637} (2002) 378,
hep-th/0204261. 

\bibitem{CKR}
B. Craps, D. Kutasov and G. Rajesh, ``String Propagation in the 
Presence of Cosmological Singularities,'' {\em JHEP} {\bf 0206}
(2002) 053, hep-th/0205101. 

\bibitem{BR}
V. Balasubramanian and S.F. Ross, ``The dual of nothing,''
hep-th/0205290.

\bibitem{KM}
S. Kachru and L. McAllister, ``Bouncing Brane Cosmologies from 
Warped String Compactifications,'' hep-th/0205209. 

\bibitem{Bir}
D. Birmingham and M. Rinaldi, ``Bubbles in 
Anti-de Sitter Space,'' hep-th/0205246.

\bibitem{Law}
A. Lawrence, ``On the instability of 3d null singularities,''
hep-th/0205288. 

\bibitem{LMS2}
H. Liu, G. Moore and N. Seiberg, ``Strings in Time-Dependent Orbifolds,''
hep-th/0206182.

\bibitem{FMc}
M. Fabinger and J. McGreevy, ``
On Smooth Time-Dependent Orbifolds and Null Singularities,''
hep-th/0206196.

\bibitem{HPo}
G.T. Horowitz and J. Polchinski, ``Instability of Spacelike and Null 
Orbifold Singularities,'' hep-th/0206228.

\bibitem{Ghez} 
A.M. Ghezelbash and R.B. Mann. ``Nutty Bubbles,''
hep-th/0207123.

\bibitem{BLW}
A. Buchel, P. Langfelder and J. Walcher, ``On Time-dependent 
Backgrounds in Supergravity and String Theory,''
hep-th/0207214. 

\bibitem{HKK}
S. Hemming, E. Keski-Vakkuri and P. Kraus, ``Strings in the 
Extended BTZ Spacetime,'' hep-th/0208003.
 
\bibitem{HS}
A. Hashimoto and S. Sethi, ``Holography and String Dynamics in 
Time-Dependent Background,'' hep-th/0208126.

\bibitem{Sim}
J. Simon, ``
Null orbifolds in AdS, Time Dependence and Holography,''
hep-th/0208165.

\bibitem{GMSS}
R. Gopakumar, S. Minwalla, N. Seiberg and A. Strominger, ``
OM Theory in Diverse Dimensions,'' {\em JHEP} {\bf 0008}
(2000) 008, hep-th/0006062.

\bibitem{BDGKR}
A. Bergman, K. Dasgupta, O. J. Ganor, J. L. Karczmarek and 
G. Rajesh, ``Nonlocal Field Theories and their Gravity duals,''
{\em Phys. Rev. } {\bf D65} (2002) 066005, hep-th/0103090.

\bibitem{AY}
M. Alishahiha and H. Yavartanoo, ``Supergravity Description of 
the Large N Noncommutative Dipole Field Theories,''
{\em JHEP} {\bf 0204} (2002) 031, hep-th/0202131.

\bibitem{T-D}
A. Giveon, M. Porrati and E. Rabinovici, ``Target Space 
Duality in String Theory,'' {\em Phys.Rept.} {\bf 244} 
(1994) 77, hep-th/9401139. 

E. Bergshoeff, C.M. Hull and T. Ortin, ``Duality in the 
Type-II Superstring Effective Action,'' {\em Nucl.Phys.}
{\bf B451} (1995) 547, hep-th/9504081. 

\bibitem{Ofa1}
J. Figueroa-O'Farrill and Joan Simon, ``Supersymmetric 
Kaluza-Klein reductions of M2 and M5-branes,'' 
hep-th/0208107.

\bibitem{Ofa2}
J. Figueroa-O'Farrill and Joan Simon, ''
Supersymmetric Kaluza-Klein reductions of M-waves 
and MKK-monopoles,'' hep-th/0208108. 


\bibitem{GKT}
S.S. Gubser, I.R. Klebanov and A.A. Tseytlin, ``String Theory and 
Classical Absorption by Threebranes'', {\em Nucl.Phys.} {\bf B499}
(1997) 217; hep-th/9703040. 

\bibitem{GK}
S. S. Gubser, I. R. Klebanov, ``Absorption by Branes and Schwinger 
Terms in the World Volume Theory'', {\em Phys.Lett.} {\bf B413} 
(1997) 41; hep-th/9708005.



\bibitem{AIO}
M. Alishahiha, H. Ita and Y. Oz, ``Graviton Scattering on D6 
Branes with B Fields'', {\em JHEP} {\bf 0006} (2000) 002; 
hep-th/0004011.

\bibitem{MAL}
J. Maldacena, ``The Large N Limit of Superconformal Field 
Theories and Supergravity ,''
{\em Adv.Theor.Math.Phys.} {\bf 2} (1998) 231, hep-th/9711200.

\bibitem{IMSY}
N. Itzhaki, J. M. Maldacena, J. Sonnenschein, S. Yankielowicz, ``Supergravity 
and The Large N Limit of Theories With Sixteen Supercharges'', {\em Phys.Rev.}
{\bf D58} (1998) 046004; hep-th/9802042. 


\bibitem{MOS}
M. Alishahiha, Y. Oz and M.M. Sheikh-Jabbari, ``Supergravity and 
Large N Noncommutative Field Theories,'' {\em JHEP} {\\be 9911} 
(1999) 007, hep-th/9909215.

\bibitem{allpapers}
T. Harmark and N.A. Obers, ``Phase Structure of Non-Commutative 
Field Theories and Spinning Brane Bound States,'' {\em JHEP}
{\bf 0003} (2000) 024, hep-th/9911169.

J.L.F. Barbon and E. Rabinovici, ``On 1/N Corrections to the
Entropy of Noncommutative Yang-Mills Theories,'' {\em JHEP}
{\bf 9912} (1999) 017, hep-th/9910019.

J. X. Lu and S. Roy, ``(p + 1)-Dimensional Noncommutative Yang-Mills
and D(p - 2) Branes,'' {\em Nucl.Phys.} {\bf B579} (2000) 229m 
hep-th/9912165.

R.-G. Cai and N. Ohta, ``Noncommutative and Ordinary Super 
Yang-Mills on (D$(p-2)$, D$p$) Bound States,'' {\em JHEP}
{\bf 0003} (2000) 009,
hep-th/0001213. 

D. S. Berman, V. L. Campos, M. Cederwall, U. Gran, 
H. Larsson, M. Nielsen, B. E.W. Nilsson and P. Sundell, 
``Holographic Noncommutativity,'' {\em JHEP} {\bf 0105}
(2001) 002, hep-th/0011282. 

\bibitem{GMMS}
R. Gopakumar, J. Maldacena, S. Minwalla and A. Strominger, ``
S-Duality and Noncommutative Gauge Theory,'' {\em JHEP}
{\bf 0006} (2000) 036, hep-th/0005048.
 
\bibitem{HARM1}
T. Harmark, ``Open Branes in Space-Time Non-Commutative 
Little String Theory,'' {\em Nucl.Phys.} {\bf B593} 
(2001) 76, hep-th/0007147. 


\bibitem{all}
T. Harmark, ``Supergravity and Space-Time Non-Commutative 
Open String Theory,'' {\em JHEP} {\bf 0007} (2000) 043,
hep-th/0006023.


\bibitem{ALI1}
M. Alishahiha, ``On Type II NS5-branes in the presence of an RR 
field,'' {\em Phys.Lett.} {\bf B486}(2000) 194, hep-th/0002198

 
\bibitem{AOR}
M. Alishahiha, Y. Oz and J. G. Russo, ``Supergravity and 
Light-Like Non-commutativity,'' {\em JHEP} {\bf 0009} (2000) 002,
hep-th/0007215. 

\bibitem{Mitra}
I. Mitra and S. Roy, ``(NS5,Dp) and (NS5,D(p+2),Dp) bound 
states of type IIB and type IIA string theories,''
{\em JHEP} {\bf 0102} (2001) 026, hep-th/0011236


\bibitem{Seiberg}
N. Seiberg, ``Matrix Description of M-theory on $T^5$ 
and $T^5/Z_2$,'' {\em Phys.Lett.} {\bf B408} (1997) 98, 
hep-th/9705221.

\bibitem{BER}
M. Berkooz, M. Rozali and N. Seiberg, ``
Matrix Description of M-theory on $T^4$ and $T^5$,''
{\em Phys.Lett.} {\bf B408} (1997) 105, hep-th/9704089.

R. Dijkgraaf, E. Verlinde and H. Verlinde, ``BPS Quantization 
of the Five-Brane,'' {\em Nucl.Phys.} {\bf B486} (1997) 89, 
hep-th/9604055.

\bibitem{MS}
S. Minwalla and N. Seiberg, ``Comments on the IIA NS5-brane,''
{\em JHEP} {\bf 9906} (1999) 007,
hep-th/9904142. 

\bibitem{DGR}
K. Dasgupta, O. J. Ganor and G. Rajesh, ``
Vector Deformations of N=4 Super-Yang-Mills Theory, Pinned 
Branes, and Arched Strings,'' {\em JHEP} {\bf 0104} (2001) 
034, hep-th/0010072.

\bibitem{Jabbari}
K. Dasgupta and M. M. Sheikh-Jabbari, ``Noncommutative Dipole 
Field Theories,'' {\em JHEP} {\bf 0202}
(2002) 002, hep-th/0112064. 


\bibitem{MAL11}
J. M. Maldacena, ``Wilson loops in Large $N$ Field theories,'' 
{\em Phys. Rev. Lett.} {\bf 80} (1998) 4859, hep-th/9803002.

\bibitem{REY}
S.-J. Rey and J. Yee. ``Macroscopic String as Heavy Quarks of
Large $N$ Gauge theory and Anti-de-Sitter Supergravity,''
{\em Eur. Phys. J.} {\bf C22} (2001) 379, hep-th/9803001. 









\end{thebibliography}
\end{document}